\documentclass{PoS}
 
\def\gsim{\;\lower4pt\hbox{${\buildrel\displaystyle >\over\sim}$}\;}
\def\lsim{\;\lower4pt\hbox{${\buildrel\displaystyle <\over\sim}$}\;}
\def\grls{\;\lower4pt\hbox{${\buildrel\displaystyle >\over <}$}\;}

\title{Mass and Mean Velocity Dispersion Relations for
Supermassive Black Holes in Galactic Bulges}

\ShortTitle{Mass and Velocity Dispersion Relations for SMBH in
Galactic Bulges}

\author{Yu-Qing Lou
\\
Physics Department and Tsinghua Center for Astrophysics, Tsinghua University,
Beijing 100084, China\\
E-mail: \email{louyq@tsinghua.edu.cn }}

\author{Yan-Fei Jiang
\thanks{Now at Department of Astrophysical Sciences, Princeton
University, Princeton, NJ 08544 USA.}\\
Physics Department and Tsinghua Center for Astrophysics, Tsinghua
University, Beijing 100084, China\\
       E-mail: \email{jiangyanfei1986@gmail.com}}

\abstract{Growing evidence indicate supermassive black holes (SMBHs)
in the mass range of $M_{\rm BH}$$\sim 10^6-10^{10}M_{\odot}$
lurking in central bulges of many galaxies
~\cite{kormendy1995,kormendy2004}. Extensive observations reveal
fairly tight power laws of $M_{\rm BH}$ versus the mean stellar
velocity dispersion $\sigma$ of the host bulge
~\cite{ferrarese2005,hu2008,tremaine2002}. The dynamic evolution of
a bulge and the formation of a central SMBH should be physically
linked by various observational clues. In this contribution, we
reproduce the empirical $M_{\rm BH}-\sigma$ power laws based on a
self-similar general polytropic quasi-static bulge evolution
~\cite{louwang2006,loujiangjin2008} and a sensible criterion of
forming a SMBH surrounding the central density singularity of a
general singular polytropic sphere (SPS)~\cite{loujiang2008}. Other
properties of host bulges and central SMBHs are also examined.
Based on our model, we discuss the intrinsic scatter of the
$M_{\rm BH}-\sigma$ relation and a scenario for the evolution of
SMBHs in different host bulges.}

\FullConference{VII Microquasar Workshop: Microquasars and Beyond\\
         September 1-5 2008\\
         Foca, Izmir, Turkey}

\begin{document}

\section*{Supermassive Black Holes and Galactic Host Bulges}

SMBHs form at the centres of elliptical and spiral galaxies
~\cite{lyndenbell1969,kormendy1995,kormendy2004}.
Observationally, SMBH masses $M_{\rm BH}$ correlate with various
properties of spiral galaxy bulges or elliptical galaxies, including
bulge luminosities
~\cite{kormendy1995,magorrianetal1998,marconihunt2003}, bulge masses
$M_{\rm
bulge}$
~\cite{magorrianetal1998,marconihunt2003,haringrix2004},
galaxy light concentrations
~\cite{graham2001}, the S$\acute{\rm e}$rsic
index of surface brightness profile
~\cite{graham2007}, inner core radii
~\cite{laueretal2007},
bulge gravitational binding energies
~\cite{aller2007} and mean stellar velocity dispersions
$\sigma$
~\cite{ferrarese2000,gebhardtetal2000,tremaine2002,ferrarese2005,hu2008}.
These correlations strongly suggest a dynamical link between SMBHs
and their host bulges
~\cite{springel2005,lietal2007}.

Among these relations, $M_{\rm BH}$ and $\sigma$ correlate tightly
in power laws with an intrinsic scatter $\lsim 0.3$ dex
~\cite{novak2006}. This relation was studied
~\cite{silkrees1998,fabian1999,blandford1999} before systematic
observations
~\cite{ferrarese2000,gebhardtetal2000} and the emphasis was on
outflow effects of galaxies. The idea was further elaborated
~\cite{king2003}. A model of singular isothermal sphere (SIS) with
rotation
~\cite{adams2001} was proposed for the $M_{\rm BH}-\sigma$
relation. This relation was also explored in a semi-analytic
model
~\cite{kauffmann2000} with starbursts while SMBHs being formed and
fueled during major mergers. Accretions of collisional dark matter
onto SMBHs may also give the $M_{\rm BH}-\sigma$ relation
~\cite{ostriker2000} (see also ref.~\cite{haehnelt2004} for a
review). There are also numerical simulations to model feedbacks
from SMBHs and stars on host galaxies.

There are two empirical types of bulges, namely classical bulges
(spiral galaxies with classical bulges or elliptical galaxies) and
pseudobulges
~\cite{
droryfisher2007}. While SMBHs in classical bulges are formed after
major mergers, pseudobulges do not show apparent merger
signatures. Interestingly, pseudobulges also manifest a $M_{\rm
BH}-\sigma$ power law yet with a different exponent
~\cite{
hu2008}.

The self-similar quasi-static solutions take the singular polytropic
spheres (the static polytropic solutions) as the leading terms. This
kind of dynamic solutions has been applied to study the compact
stars in a single fluid model~\cite{louwang2006}, galaxy clusters in
a two-fluid model~\cite{loujiangjin2008} and the so-called
``champagne flows" in H II regions~\cite{hulou2008}. In the
following, we shall take the quasi-static solutions in a single
fluid model to describe host bulges with central SMBHs after long
time evolution~\cite{loujiang2008}.

\section*{A Self-Similar Dynamic Model for $M_{\rm BH}-\sigma$ Power Laws}

For the dynamic evolution of a galactic bulge, we adopt a few
assumptions. First, we treat the stellar bulge as a spherical
polytropic fluid as the typical age $\sim10^9$ yr of galactic bulges
is long
~\cite{frogel1988,gnedin1999} that they are continuously adjusted.
Stellar velocity dispersions produce an effective pressure $P$
against the self-gravity as in the Jeans equation
~\cite{binneytremaine1994}. Secondly,
the total mass of the interstellar medium
in a galaxy~\cite{gnedin1999}
is $\sim 10^7-10^8 M_{\odot}$
only $10^{-2}\sim 10^{-3}$ of the total bulge mass.
Although gas densities in broad and narrow line regions of AGNs
are high, the filling factor~\cite{osterbrock2006}
is small
($\sim 10^{-3}$)
and the gas there may be regarded as condensed clouds. Thus gas is
merged into our stellar fluid. Thirdly, the diameter of broad line
regions of AGNs~\cite{osterbrock2006}
is only $\sim 0.1 $ pc
and the disk around a SMBH is even smaller while a galactic bulge
size is $\gsim 1$ kpc. We thus ignore small-scale structures around
the central SMBH of a spherical bulge. Finally, as the rotation
curves of galaxies show~\cite{binneytremaine1994}, the effect of
dark matter halo in the innermost region (around several kpcs ) of a
galaxy can be neglected. So we do not include the dark matter in our
model when we discuss the dynamic characters of the bulges.

Hydrodynamic equations of our model are conservations of mass,
radial momentum~\cite{
louwang2006}
and `specific entropy' along streamlines
~\cite{
wanglou2007,loucao2008,loujiangjin2008}. As bulk flow of stellar
fluid is slow, we invoke the novel self-similar quasi-static
solutions
~\cite{louwang2006,loujiangjin2008} to model the bulge evolution.
We use a self-similar transformation
~\cite{
louwang2006,loucao2008,loujiangjin2008} to solve general
polytropic fluid equations, namely
\begin{eqnarray}
r=K^{1/2}t^{n}x\ ,\quad\rho={\alpha(x)}/{(4\pi Gt^2)}\
,\quad u=K^{1/2}t^{n-1}v(x)\ ,\nonumber\\
P={Kt^{2n-4}\beta(x)}/{(4\pi G)}\ ,\qquad
M={K^{3/2}t^{3n-2}m(x)}/{[(3n-2)G]}\ ,\label{transformation}
\end{eqnarray}
where $r$ is the radius and $t$ is time; $x$ is the independent
dimensionless similarity variable while $K$ and $n$
are two scaling indices; $G=6.67\times10^{-8}\
\hbox{g}^{-1}\hbox{cm}^3\hbox{s}^{-2}$ is the gravity constant;
$\rho(r,t)$ is the mass density and $\alpha(x)$ is the reduced mass
density; $u(r,\ t)$ is the radial flow speed and $v(x)$ is the
reduced flow speed; $P(r,\ t)$ is the effective pressure and
$\beta(x)$ is the reduced pressure; $M(r,\ t)$ is the enclosed mass
and $m(x)$ is the reduced enclosed mass; reduced variables $\alpha,\
\beta,\ v$ and $m$ are functions of $x$ only. We require $n>2/3$ for
a positive mass
~\cite{louwang2006,loujiangjin2008,loucao2008}.

By self-similar transformation (\ref{transformation}), we readily
construct self-similar quasi-static solutions from general
polytropic fluid equations, taking the static singular polytropic
sphere (SPS) as the leading term. Properties of such asymptotic
solutions to the leading order
~\cite{louwang2006,loujiangjin2008} are summarized below. Both
initial and eventual
mass density profiles scale as $\sim r^{-2/n}$;
accordingly, the bulge enclosed mass profile is $M\propto
r^{(3n-2)/n}$; for either $x\rightarrow 0^{+}$ or $x\rightarrow
+\infty$, the reduced velocity $v\rightarrow 0$, which means at a
time $t$, for either $r\rightarrow0^+$ or $r\rightarrow+\infty$
the flow speed $u\rightarrow 0$, or at a radius $r$, when $t$ is
short or long enough, the radial flow speed $u\rightarrow 0$. Our
model~\cite{loujiang2008} describes a self-similar bulge evolution
towards a nearly static configuration after a long time lapse,
appropriate for galactic bulges at present epoch.

As the effective pressure $P$ results from stellar velocity
dispersions in the bulge, we readily derive the mean velocity
dispersion $\sigma$ in a bulge. By specific entropy conservation
along streamlines, we relate $P$ with $\rho$ and $M$
and derive the $P$ profile from our quasi-static solutions
~\cite{wanglou2007,loujiangjin2008}.
We take the local stellar velocity dispersion as $\sigma_L(r,\
t)=(\gamma P/\rho)^{1/2}$ where $\gamma$ is the polytropic index
of our stellar fluid. To compare with observations, we derive the
spatially averaged stellar velocity dispersion $\sigma$ in the
bulge.
The bulge boundary is taken as the radius $r_{\rm c}$ where $\rho$
drops to a value $\rho_c$ indistinguishable from the environment.
For a class of bulges with same $n$, $\rho_c$ is regarded as a
constant for a class of environments.
One can show that within $r_c$
\begin{eqnarray}
\sigma=[3n^{1+q/2}\gamma^{1/2}/(4n-1)](4\pi
G\rho_c)^{(1-n)/2}A^{3nq/4}K^{1/2}\equiv {\cal Q}K^{1/2}\
,\label{sigma}
\end{eqnarray}
where $q\equiv 2(n+\gamma-2)/(3n-2)$ and $A\equiv
\{n^{2-q}/[2(2-n)(3n-2)]\}^{-1/(n-3nq/2)}$.

\begin{figure}
\begin{center}
  \includegraphics[height=7cm, width=9cm]{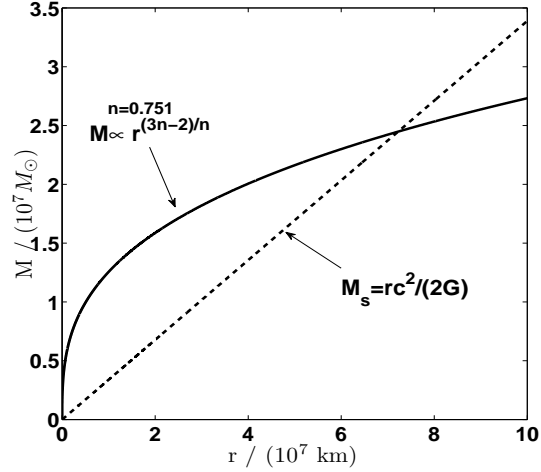}
\end{center}
\caption {\footnotesize The criterion of forming a central SMBH in a
self-similar quasi-static bulge evolution. The enclosed mass
power-law is $M\propto r^{0.337}$ with $n=0.751$ (solid curve).
Meanwhile, we draw a straight dashed line $M_{\rm s}=rc^2/(2G)$ for
the mass of SMBH versus the Schwarzschild radius $r$. Here at
$r_{\rm s}=7.2\times10^7$ km, the straight line intersects the
enclosed mass curve for a $2.45\times 10^7M_{\odot}$ SMBH.
A self-similar general polytropic quasi-static solution of $n<1$ can
form a SMBH at the bulge centre by invoking this
criterion.}\label{mass}
\end{figure}

A SMBH forms at the centre of a galactic bulge that evolves in a
self-similar quasi-static manner. Such a SMBH was formed by the
collapse of collections of stars and gas towards the bulge centre
and grows rapidly by matter accretions at an earlier phase
~\cite{
lyndenbell1969}.
As the growth timescale for SMBHs is only $\sim 10^5$ yr, our
quasi-static solutions describe the relatively quiescent phase of
galactic bulges after the formation of central SMBHs as a longer
history of a bulge evolution. The stellar fluid made up of stars and
condensed gas clouds has a slow bulk flow speed towards the central
SMBH, sustaining a reservoir of mass accretion for the circumnuclear
torus and/or disk.
We now introduce the criterion of forming a SMBH. A SMBH mass
$M_{\rm BH}$ and
its Schwarzschild radius $r_{\rm s}$ are related by $M_{\rm
BH}=r_{\rm s}c^2/(2G)$ where $c$
is the speed of light. As we have pointed out, the bulge enclosed
mass is $M\propto r^{(3n-2)/n}$. If at a radius $r_{\rm s}$, $M$
becomes
$M=r_{\rm s}c^2/(2G)$, then a SMBH forms.
Only those quasi-static bulges with $n<1$ can form central SMBHs
(see Figure \ref{mass});
we further derive
$M_{\rm BH}\propto r_{\rm s}\propto K^{1/(2-2n)}$ and the power law
below
\begin{eqnarray}
M_{\rm BH}= \Bigg(\frac{nA}{3n-2}\Bigg)^{n/(2-2n)}
\Bigg(\frac{2}{c^2}\Bigg)^{3n-2/(2-2n)}\frac{{\cal Q}^{1/(n-1)}}{G}\
\sigma^{\ 1/(1-n)}\equiv\mathcal{L}\ \sigma^{\ 1/(1-n)}\
,\label{relation}
\end{eqnarray}
where $\mathcal{L}$ depends on $c$, $G$, $n$, $\gamma$, $\rho_c$,
and the exponent $1/(1-n)>3$ as $2/3<n<1$ is required.

\begin{figure}
\begin{center}
  \includegraphics[height=7cm,width=9cm]{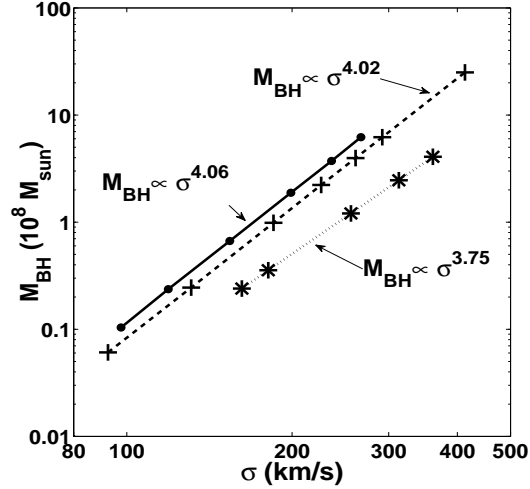}
\end{center}
\caption{
Power-law $M_{\rm BH}-\sigma$ relations according to our general
polytropic model for quasi-static self-similar evolution. Three
index values $n=0.7537$, $0.7512$, $0.7333$ are adopted for three
different kinds (solid
~\cite{hu2008}, dashed
~\cite{tremaine2002}, dotted lines
~\cite{gebhardtetal2000}) of $M_{\rm BH}-\sigma$ power laws given by
equation $(3)$.
Given $n$, we calculate the SMBH mass $M_{\rm BH}$ and the mean
stellar velocity dispersion $\sigma$ for a certain $K$ value in
transformation. A bulge has different $\sigma$ for different $K$.
Each bulge-SMBH system is represented by a point in this display.
All systems of same $n$ give a straight line, while systems of
different $n$ correspond to different straight lines
here.}\label{msigmarelation}
\end{figure}

While the $M_{\rm BH}-\sigma$ power law is very tight with intrinsic
scatter $\le 0.3$ dex for SMBHs and host bulges
~\cite{novak2006}, such scatter is large enough for different
fitting parameters
~\cite{ferrarese2000,gebhardtetal2000,tremaine2002}. By equation
(\ref{relation}), we have a natural interpretation for intrinsic
scatter in the observed $M_{\rm BH}-\sigma$ power law. In our
model, all bulges with the same $n$ lie on a straight line with
the exponent $1/(1-n)$ as Figure \ref{msigmarelation} shows. For a
fixed $n$, different bulges are represented by different $K$
values in transformation (\ref{transformation}), leading to
different $M_{\rm BH}$ and $\sigma$. However, for bulges with
different $n$ values, they lie on different lines. For elliptical
galaxies or bulges in spiral galaxies, they appear to eventually
take the self-similar evolution described above with a certain $n$
value. But pseudobulges may take on different $n$ values.
Observationally, we cannot determine a priori the specific $n$
value for a bulge but simply attempt to fit all bulges with a
single exponent, which then contributes in part to intrinsic
scatter in the observed $M_{\rm BH}-\sigma$ power law.

To show this, we fit three published $M_{\rm BH}-\sigma$ power laws
in Figure \ref{msigmarelation}. The first one is $M_{\rm
BH}=1.2\times10^8M_{\odot}(\sigma/200\hbox{ km s}^{-1})^{3.75}$
given in ref.~\cite{gebhardtetal2000} with our parameters $\{n,\
\gamma,\ \rho_c\}$ being $\{0.733,\ 1.327,\ 0.47\ M_{\odot}
\hbox{pc}^{-3}\}$; and the five points correspond to $K=\{0.8,\ 1,\
2,\ 3,\ 4 \}\times10^{23}$ cgs unit. The second one is $\log (M_{\rm
BH}/M_{\odot})=8.13+4.02\log (\sigma/ 200\hbox{ km s}^{-1})$ given
in ref.~\cite{tremaine2002} with our parameters $\{n,\ \gamma,\
\rho_c\}$ being $\{0.7512$, 1.33, $0.0122\
M_{\odot}\hbox{pc}^{-3}\}$; and the seven points correspond to
$K=\{1,\ 2,\ 4,\ 6,\ 8,\ 10,\ 20 \}\times10^{22}$ cgs unit. The
third one is $\log (M_{\rm BH}/M_{\odot})=8.28+4.06\log (\sigma/
200\hbox{ km s}^{-1})$ given in ref.~\cite{hu2008} with our
parameters $\{n,\ \gamma,\ \rho_c\}$ being $\{0.7537,\ 1.332$,
$0.00364\ M_{\odot} \hbox{pc}^{-3}\}$; and the six points correspond
to $K=\{0.6,\ 0.9,\ 1.5,\ 2.5,\ 3.5,\ 4.5 \}\times10^{23}$ cgs unit.
Clearly, to fit all these points in Figure \ref{msigmarelation} with
a single power law, we would get a different result with higher
intrinsic scatter. In the three fitting examples, bulge inflow
speeds of stellar fluid are slow ($\sim 0.1-1$ km s$^{-1}$), a
feature of our self-similar quasi-static solutions. Near the SMBH
boundary $r_{\rm s}$, the inflow rest mass-energy flux falls in the
range of $10^{40}-10^{45}\hbox{ erg s}^{-1}$ in these examples,
sufficient to supply the observed X-ray luminosities
~\cite{komossa2008}. There can be outgoing accretion shocks around
a SMBH in these flows. As the age of galactic bulges
~\cite{frogel1988} is so long ($\sim10^{9}$ yr) that such shocks
should have gone outside bulges.

Besides the $M_{\rm BH}-\sigma$ relation,
observations reveal $M_{\rm BH}\propto M_{\rm bulge}^{1.12}$ with
$M_{\rm bulge}$ being the bulge mass
~\cite{haringrix2004} and $M_{\rm BH}\propto E_{\rm g}^{0.6}$ with
$E_{\rm g}$ being the absolute value of the bulge gravitational
binding energy
~\cite{aller2007}. Using our criterion of forming a SMBH and the
bulge radius $r_c$, we derive a power law between $M_{\rm BH}$ and
$M_{\rm bulge}$ as $M_{\rm BH}\propto M_{\rm bulge}^{1/(3-3n)}$.
For $n=0.75$, our result leads to relations in
ref.~\cite{adams2001} but for a nonisothermal general SPS. The
bulge gravitational binding energy, without contributions from
dark matter halo and a disk
~\cite{aller2007}, is $E_{\rm
g}\approx\int_0^{r_c}{GM}\rho 4\pi rdr$. For self-similar
quasi-static solutions, we obtain $M_{\rm BH}\propto E_{\rm
g}^{1/(5-5n)}$.

As another class of bulges, pseudobulges are believed to have
formed without merging in contrast to classical bulges.
Nonetheless, pseudobulges follow a $M_{\rm BH}-\sigma$ power law
but with a different exponent
~\cite{kormendy2001,hu2008}.
In ref.~\cite{hu2008}, this $M_{\rm BH}-\sigma$ relation is found
to be $\log (M_{\rm BH}/M_{\odot})=7.5+4.5\log (\sigma/ 200\hbox{
km s}^{-1})$. This conclusion is natural according to our model in
that pseudobulges may take a different self-similar quasi-static
evolution for a different $n$ value. Due to their different
formation history, they show a different $M_{\rm BH}-\sigma$ power
law as observed. For $\{n,\ \gamma,\ \rho_c\}$ being $\{0.7778,\
1.34,\ 0.000426\ M_{\odot}\hbox{pc}^{-3}\}$, our model can also
fit this power law.

\section*{Conclusions and Discussion}

The tight $M_{\rm BH}-\sigma$ power laws and other relations among
the SMBH mass $M_{\rm BH}$ and known properties of host bulges
strongly suggest coeval growths of SMBHs and galactic
bulges
~\cite{pageetal2001,haehnelt2004,kauffmann2000}. In our model,
while forming a Schwarzschild SMBH at the bulge centre (e.g., by
collapse of gas and stars or by merging), the spherical general
polytropic bulge evolves in a self-similar quasi-static phase for
a long time.
We then reproduce empirical $M_{\rm BH}-\sigma$ power laws.
Different energetic processes appear to give rise to different
scaling index $n$ values.

Besides classical bulges and pseudobulges, there are also `core'
elliptical galaxies, thought to have formed by `dry' mergers. A
steeper $M_{\rm BH}-\sigma$ relation exists in these galaxies as
compared to that for classical bulges
~\cite{laueretal2007}.
This can be accommodated in our scenario that all hosts of SMBHs may
finally evolve into self-similar quasi-static phase with different
scaling parameters (e.g., different $n$ for the slope and different
$\rho_c$ for the normalization of the $M_{\rm BH}-\sigma$ relation).
We thus provide a unified framework to model the the relatively
quiescent evolution phase of SMBH host bulges and SMBH masses. As
the observed $M_{\rm BH}-\sigma$ relation for classical bulges is
tight, the elliptical galaxies and spiral galaxies appear to take
close $n$ values for merging processes.

In our model, $n$ is a key scaling index to determine the exponent
of the $M_{\rm BH}-\sigma$ power law. The smaller the value of $n$
is, the steeper the profile of the density is and the smaller the
index of the $M_{\rm BH}-\sigma$ relation is. If we think the SMBHs
are formed by collapse of stars and gas and a less steeper density
distribution may provide a more effective mechanism to form SMBHs,
then we conclude that for a certain value of velocity dispersions,
the smaller the mass of the initially formed SMBH is, the smaller
the value of $n$ is.




\section*{Acknowledgement}
This research was supported in part
by Tsinghua Center for Astrophysics (THCA), National Science
Foundation of China (NSFC) grants 10373009 and 10533020, the Yangtze
Endowment and SRFDP 20050003088 of Ministry of Education, and NBSTTP
J0630317 of NSFC at Tsinghua University. The kind hospitality of
Institut f\"ur Theoretishce Physik und Astrophysik der
Christian-Albrechts-Universit\"at Kiel is gratefully acknowledged.


\begin{thebibliography}{99}

\bibitem{kormendy1995}
Kormendy, J. \& Richstone, D. Inward bound - the search for
supermassive black holes in galactic nuclei. {\it Annu. Rev. Astron.
Astrophys.} {\bf 33}, 581-624 (1995)

\bibitem{kormendy2004}
Kormendy, J. Coevolution of black holes and galaxies. {\it Obs.
Astrophys. Ser.} {\bf 1}, 1 (2004)

\bibitem{ferrarese2005}
Ferrarese, L. \& Ford, H. Supermassive blackholes in galactic
nuclei: past, present and future research. {\it Space Sci. Rev.}
{\bf 116}, 523-624 (2005)

\bibitem{hu2008}
Hu, J. The black hole mass-stellar velocity dispersion correlation:
bulges versus pseudo-bulges. {\it Mon. Not. Roy. Astron. Soc.} in
press (2008), (arxiv: 0801.1481v3)

\bibitem{tremaine2002}
Tremaine, S. {\it et al.} The slope of the black hole mass versus
velocity dispersion correlation. {\it Astrophys. J.} {\bf 574},
740-753 (2002)

\bibitem{louwang2006}
Lou, Y.-Q. \& Wang, W. G. New self-similar solutions of polytropic
gas dynamics. {\it Mon. Not. Roy. Astron. Soc.} {\bf 372}, 885-900
(2006)

\bibitem{loujiangjin2008}
Lou, Y.-Q., Jiang, Y. F. \& Jin, C. C. Self-similar shocks and winds
in galaxy clusters. {\it Mon. Not. Roy. Astron. Soc.} {\bf 386},
835-858 (2008)

\bibitem{loujiang2008}
Lou, Y.-Q. \& Jiang, Y. F. Supermassive black holes in galactic
bulges. {\it Mon. Not. Roy. Astron. Soc. Letter}, in press (2008)
(2008arXiv:0809.1126L)

\bibitem{hulou2008}
Hu, R. Y. \& Lou, Y.-Q.  Self-similar polytropic champagne flows in
H II regions. {\it Mon. Not. Roy. Astron. Soc.}, in press (2008)
(2008arXiv0808.2090H)

\bibitem{lyndenbell1969}
Lynden-Bell, D. Galactic nuclei as collapsed old quasars. {\it
Nature} {\bf 223}, 690-695 (1969)

\bibitem{marconihunt2003}
Marconi, A. \& Hunt, L. K. The relation between black hole mass,
bulge mass, and near-infrared luminosity $M_{\rm BH}-\sigma$ and
mass relations. {\it Astrophys. J.} {\bf 589}, L21-24 (2003)

\bibitem{magorrianetal1998}
Magorrian, J. {\it et al.} The demography of massive dark objects in
galaxy centres. {\it Astron. J.} {\bf 115}, 2285-2312 (1998)

\bibitem{haringrix2004}
H$\ddot{{\rm a}}$ring, N. \& Rix, H.-W. On the black hole
mass¨cbulge mass relation. {\it Astrophys. J.} {\bf 604}, L89-92
(2004)

\bibitem{graham2001}
Graham, A. W., Erwin P., Caon N. \& Trujillo I.,
A correlation between galaxy light
concentration and supermassive black hole mass. {\it Astrophys. J.}
{\bf 563}, L11-14 (2001)

\bibitem{graham2007}
Graham, A. W. \& Driver, S. P. A log-quadratic relation for
predicting supermassive black hole masses from the host bulge sersic
index. {\it Astrophys. J.} {\bf 655}, 77 (2007)

\bibitem{laueretal2007}
Lauer, T. R. {\it et al.}
Masses of nuclear black holes in luminous elliptical galaxies and
implications for the space density of the
massive black holes. {\it Astrophys. J.} {\bf 662}, 808-834 (2007)

\bibitem{aller2007}
Aller, M. C. \& Richstone, D. O. Host galaxy bulge predictors of
supermassive black hole mass. {\it Astrophys. J.} {\bf 665}, 120-156
(2007)

\bibitem{ferrarese2000}
Ferrarese, L. \& Merritt, D. A fundamental relation between
supermassive black holes and their host galaxies. {\it Astrophys.
J.} {\bf 539}, L9-12 (2000)

\bibitem{gebhardtetal2000}
Gebhardt, K. {\it et al.} A relationship between nuclear black hole
mass and galaxy velocity dispersion. {\it Astrophys. J.} {\bf 539},
L13-16 (2000)

\bibitem{lietal2007}
Li, Y., Haiman, Z. \& Low,  M. M. Correlations between central
massive objects and their host galaxies: from bulgeless spirals to
ellipticals. {\it Astrophys. J.} {\bf 663}, 61-70 (2007)

\bibitem{springel2005}
Springel, V., Matteo, T. D. \& Hernquist, L. Modelling feedback
fromstars and black holes ingalaxy mergers. {\it Mon. Not. Roy.
Astron. Soc.} {\bf 361}, 776-794 (2005)

\bibitem{blandford1999}
Blandford, R. D. Origin and evolution of massive black holes in
galactic nuclei. {\it Galaxy Dynamics, ASP Conf. Ser.} {\bf 182},
87-95 (1999)

\bibitem{fabian1999}
Fabian, A. C. The obscured growth of massive black holes. {\it Mon.
Not. Roy. Astron. Soc.} {\bf 308}, L39-43 (1999)

\bibitem{silkrees1998}
Silk, J. \& Rees, M. J. Quasars and galaxy formation. {\it Astron.
Astrophys.} {\bf 331}, L1-L4 (1998)

\bibitem{king2003}
King, A. Black holes, galaxy formation, and the $M_{\rm bh}-\sigma$
relation. {\it Astrophys. J.} {\bf 596}, L27-29 (2003)

\bibitem{adams2001}
Adams, F. C., Graff, D. S. \& Richstone, D. O. A theoretical model
for the $M_{\rm bh}-\sigma$ relation for supermassive black holes in
galaxies. {\it Astrophys. J.} {\bf 551}, L31-35 (2001)

\bibitem{kauffmann2000}
Kauffmann, G. \& Haehnelt, M. A unified model for the evolution of
the galaxies and quasars. {\it Mon. Not. Roy. Astron. Soc.} {\bf
311}, 576-588 (2000)


\bibitem{ostriker2000}
Ostriker, J. P. Collisional dark matter and the origin of massive
black holes. {\it Phys. Rev. Lett.} {\bf 84}, 5258-5260 (2000)

\bibitem{haehnelt2004}
Haehnelt, M. G. {\it Coevolution of Black Holes and Galaxies, from
the Carnegie Observatories Centennial Symposia}, 405 (Cambridge
Univ. Press, 2004)

\bibitem{droryfisher2007}
Drory, N. \& Fisher, D. B. A connection between bulge properties and
the bimodality of galaxies. {\it Astrophys. J.} {\bf 664}, 640-649
(2007)

\bibitem{frogel1988}
Frogel, J. A. The galactic nuclear bulge and the stellar content of
spheroidal systems. {\it Ann. Rev. Astron. Astrophys.} {\bf 26},
51-92 (1988)

\bibitem{gnedin1999}
Gnedin, N. Y., Norman, M. L. \& Ostriker, J. P. Formation of
galactic bulges. {\it AIPC} {\bf 470}, 48-57

\bibitem{binneytremaine1994}
Binney, J. \& Tremaine, S. {\it Galactic Dynamics} (Princeton
University Press, 1994)

\bibitem{osterbrock2006}
Osterbrock, D. E. \& Ferland, G. J. {\it Astrophysics of Gaseous
Nebulae and Active Galactic Nuclei} (University Science Books, 2rd
Edition, 2006)

\bibitem{novak2006}
Novak, G. S., Faber, S. M. \& Dekel, A. On the correlations of
massive black holes with their host galaxies. {\it Astrophys. J.}
{\bf 637}, 96-103 (2006)






\bibitem{komossa2008}
Komossa, S. {\it et al.} Discovery of superstrong, fading, iron line
emission and double-peaked balmer lines of the galaxy SDSS
J095209.56214313.3: The light echo of a huge flare. {\it Astrophys.
J.} {\bf 678}, L13-16 (2008)

\bibitem{kormendy2001}
Kormendy, J. \& Gebhardt, K.  {\it The 20th Texas Symposium on
Relativistic Astrophysics}, 363 (2001) in H. Martel, J.C. Wheeler,
eds.




\bibitem{loucao2008}
Lou, Y.-Q. \& Cao, Y.  Self-similar dynamics of a relativistically
hot gas. {\it Mon. Not. Roy. Astron. Soc.} {\bf 384}, 611 (2008)

\bibitem{pageetal2001}
Page, M. J., Stevens, J. A., Mittaz, J. P. D. \& Carrera, F. J.
Submillimeter evidence for the coeval
growth of massive black holes and galaxy bulges. {\it Science} {\bf
294}, 2516-2518 (2001)






\bibitem{wanglou2007}
Wang, W. G. \& Lou, Y.-Q.  Self-similar dynamics of a magnetized
polytropic gas. {\it Astropys. Space Sci.} {\bf 311}, 363 (2007)

\bibitem{wanglou2008}
Wang, W. G. \& Lou, Y.-Q.  Dynamic evolution of a quasi-spherical
general polytropic magnetofluid with self-gravity. {\it Astropys.
Space Sci.} {\bf 315}, 135 (2008)


\end{thebibliography}
\end{document}